\documentclass[a4paper,12pt]{article}
\usepackage[super,square]{natbib}
\usepackage{anysize}
\marginsize{2cm}{2cm}{1cm}{2cm}
\usepackage{fancyhdr}

\usepackage{soul}
\usepackage[dvipsnames]{xcolor}
\usepackage{hyperref}
\usepackage{graphicx}
\usepackage{tikz}
\usepackage{forest}

\hypersetup{
    colorlinks=true,
    linkcolor=black,
    citecolor=CadetBlue,
    filecolor=CadetBlue,      
    urlcolor=CadetBlue,
}
\usepackage{lipsum}
\usepackage{multicol}
\renewenvironment{abstract}
 {\par\noindent\textbf{\abstractname}\ \ignorespaces \\}
 {\par\noindent\medskip}

\begin{document}
% Makes header
\pagestyle{fancy}
\thispagestyle{empty}
\fancyhead[R]{\textit{Private Proof of Solvency}}
\fancyhead[L]{}
% Makes footnotes with an asterisk
\renewcommand*{\thefootnote}{\fnsymbol{footnote}}
\begin{center}
\Large{\textbf{Private Proof of Solvency}}
\vspace{0.4cm}
\normalsize
\\ \small{Hamid Bateni (hamid@europe.com), Keyvan Kambakhsh (keyvankambakhsh@gmail.com)}\\
\vspace{0.1cm}
\textit{Nobitex Labs}
\medskip
\normalsize
\end{center}
{\color{gray}\hrule}
\vspace{0.4cm}
\begin{abstract}
The "Private Proof of Solvency" project is a groundbreaking solution in the realm of Proof of Solvency, offering a secure, efficient, and privacy-preserving method for crypto custody providers such as centralized cryptocurrency exchanges or enterprise custody providers. By leveraging the inherent state concept of every blockchain and pioneering cryptographic techniques, our approach ensures businesses can \textbf{prove their reserves without revealing their transactions, addresses, or the total amount of liabilities}.
\end{abstract}
{\color{gray}\hrule}
\medskip
\begin{multicols}{2}
\tableofcontents
\section{Introduction}
Crypto custody providers currently face the challenge of maintaining numerous addresses for user assets. Conventional methods to create a proof of reserve require the consolidation of these assets into single or multiple known wallet addresses. Our innovative approach eliminates this process by utilizing the inherent state concept of every blockchain.

The state, achieved by processing blockchain transactions on the blockchain protocol nodes, holds data such as the balance associated with an address. For instance, Ethereum maintains this state in the Merkle Patricia data structure, while Bitcoin employs a LevelDB database with a key-value structure that keeps the active Unspent Transaction Outputs (UTXOs). In Bitcoin terms, the balance represents the total active UTXOs an address holds.

Our project introduces a novel process for businesses to provide proof of reserve:

\begin{enumerate}
    \item Create a proof of liabilities tree based on user data on the business database.
    \item Sign a message with the private key of the addresses they want to prove reserve with.
    \item Provide these messages as private input for our Zero-Knowledge Proof (ZKP) circuit.
    \item Submit the output to a contract and announce their new submission for checking.
\end{enumerate}
By leveraging ZKP, businesses can prove their reserves without the need to reveal their transactions, addresses, or the total amount of liabilities, thereby maintaining privacy while ensuring the integrity of the process. In essence, the "Private Proof of Solvency" project offers a robust, privacy-preserving solution that significantly enhances the Proof of Solvency process for crypto custody providers, paving the way for a \textbf{more secure financial ecosystem.}
\end{multicols}
{\color{gray}\hrule}
\begin{center}
\section{Proof of Liability}
\small{The first step in the proof of solvency process involves the proof of liabilities. This step aims to demonstrate the total amount of liabilities, or obligations, that exist. Liabilities in this context refer to the balances that the custody provider owes to its customers}
\cite{POS}
\end{center}
{\color{gray}\hrule}
\begin{multicols}{2}

\subsection{Commitment}
A commitment in the field of cryptography refers to a binding agreement to a chosen piece of information. Once this agreement is made, it becomes irreversible and unalterable. Essentially, it's akin to sealing a message in an envelope - the message cannot be changed once it is sealed.

In cryptographic terms, a commitment scheme allows an entity to commit to a chosen value, while keeping it hidden from others. It's designed to be both binding and hiding. Binding ensures that once the commitment is made, it cannot be changed. Hiding ensures that until the reveal, no information about the committed value is leaked.

For the Proof of Solvency process, we use a cryptographic commitment to demonstrate the existence and integrity of the liabilities. This is where we introduce the Merkle Tree as our commitment scheme. The root of the Merkle Tree serves as the commitment to all liabilities, and each leaf node of the tree represents an individual liability.

The Merkle Tree is an efficient and secure method for verifying liabilities. Using this structure, we can provide proof of the existence and integrity of liabilities without revealing the actual liabilities until necessary. This approach strikes a balance between transparency and privacy, which is crucial for crypto custody providers. In the following sections, we will discuss the structure and properties of the Merkle Tree in more detail.

\subsection{Merkle Tree}
A Merkle Tree is a tree in computer science in which every leaf node is labeled with the cryptographic hash of a data block, and each non-leaf node is labeled with the cryptographic hash of the labels of its child nodes. The value of a non-leaf node is determined by the hash of its children nodes, and this continues recursively until the tree's root is achieved.

This structure is particularly effective because it allows for efficient and secure verification of the contents of large data structures. The Merkle Tree allows us to verify data with a significantly smaller subset of the total information. \cite{hashtree}

\includegraphics[width=\linewidth]{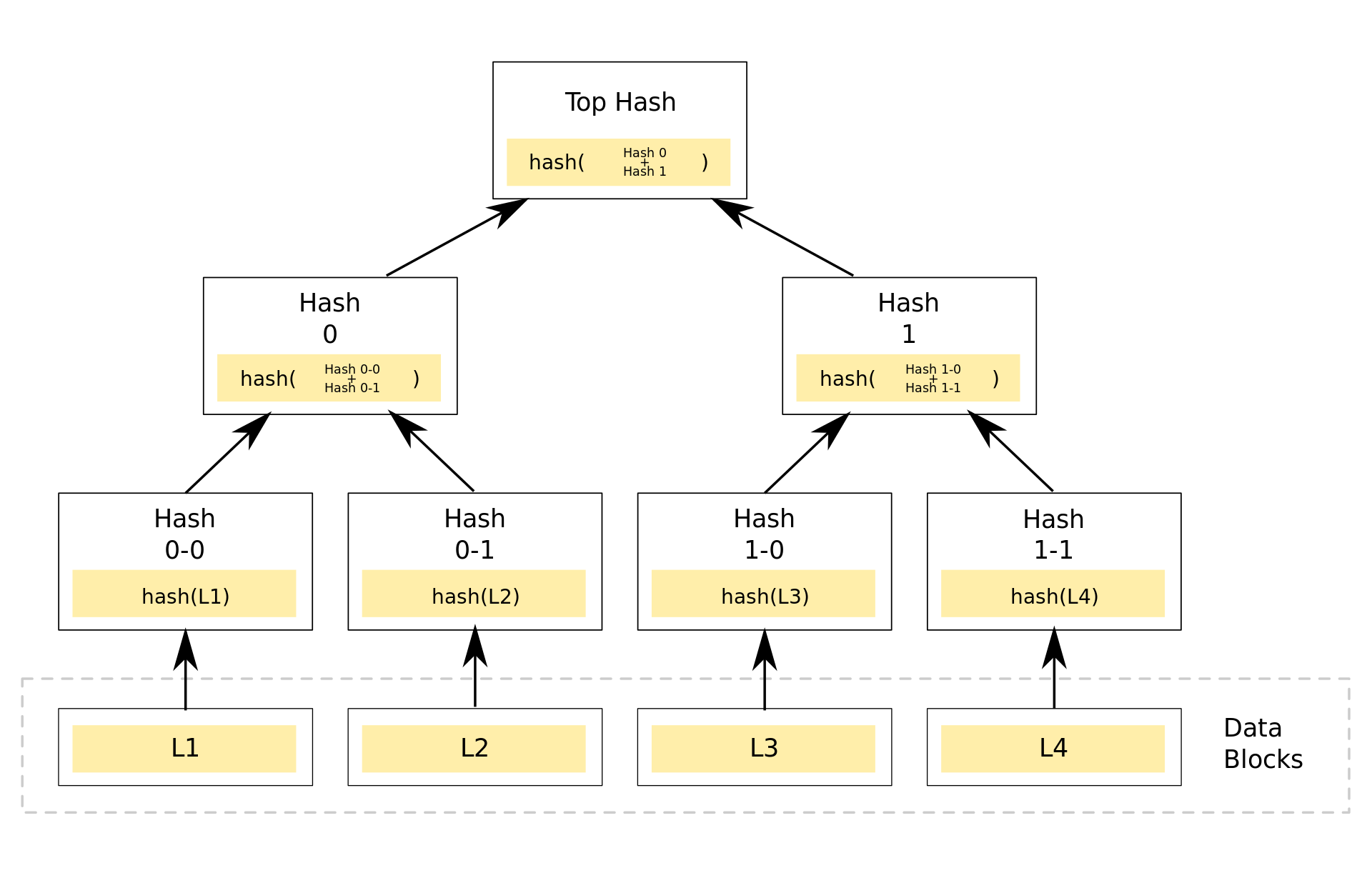}

In the scenario of Proof of Solvency, we use the Merkle Tree as our cryptographic commitment. The root of the Merkle Tree acts as a commitment to all liabilities, and each leaf in the tree corresponds to a liability. This allows for an efficient and secure way to prove the existence and integrity of liabilities.

By using a Merkle Tree, we can provide a proof path for any given leaf node (liability) up to the root. This path, also known as the Merkle path, allows anyone to verify that a specific liability is part of the tree. They can do this by recomputing the hashes from the leaf up to the root and comparing it with the root hash. This approach is particularly useful because it allows for verification without needing access to all data points, providing a balance between transparency and data efficiency.

In the following sections, we will delve deeper into the leaf structure of the Merkle Tree and how we can use it to create proof statements.

\subsection{Leaves Structure}
Until now, we have discussed how we can commit to a list of data by using a hash tree, specifically, a Merkle tree. Now let's delve into the structure of the data we are committing to.

In our commitment to data via a Merkle tree, each leaf (except the last right leaf) contains key-value pairs of data:
\newline

\setlength{\fboxsep}{10pt} % Change this to adjust the space
\noindent\fbox{%
\begin{minipage}{0.8\linewidth}
\textbf{User 1 Leaf} % This is the title
\begin{itemize}
\item User Identifier
\item Network Identifire: 
\begin{itemize}
\item Asset 1: amount
\item Asset 2: amount
\end{itemize}
\item Network Identifier: 
\begin{itemize}
\item Asset 3: amount
\item Asset 4: amount
\end{itemize}
\end{itemize}
\end{minipage}
}

\vspace{10pt} % This adds space below the box

The 'User Identifier' is a unique identifier for each user. The 'Asset Identifier' refers to the identifier for each asset. For example, in the Ethereum network, we consider the zero address as the identifier for ETH, and for each token, we consider their token address as the identifier. The 'amount' is the balance the user holds of that specific asset according to the business database that wants to prove their liabilities amount.

This approach to committing to our liabilities is similar to other proof of solvency approaches. However, traditionally, committing to our liabilities in this way would require us to make our total liabilities public. To prove the correctness of the sum, we would need to enlist the help of an auditor firm or use Zero-Knowledge Proofs (ZKP). But in doing so, we would have to reveal our total liabilities.

The approach we're introducing allows the total liabilities to remain private. Let's explain how this is achieved by discussing the last right leaf of our Merkle tree:\newline

\setlength{\fboxsep}{10pt} % Change this to adjust the space
\noindent\fbox{%
\begin{minipage}{0.8\linewidth}
\textbf{Last Right Leaf:} % This is the title
\begin{itemize}
\item Network Identifier: 
\begin{itemize}
\item Asset 1 ID: Total Balance\newline user1 + ... + user10000
\item Asset 2 ID:\newline user1 + ... + user10000
\end{itemize}
\item Network Identifier: 
\begin{itemize}
\item Asset 3 ID:\newline user1 + ... + user10000
\item Asset 4 ID:\newline user1 + ... + user10000
\end{itemize}
\end{itemize}
\end{minipage}
}
\vspace{10pt} % This adds space below the box

The Total Balance is the sum of the balances of all users for a specific asset (Asset1: user1 balance + user2 balance + ...).

We then use a ZKP protocol to create a circuit and leverage the characteristic of verifiable computation to prove that the total sum in the last right leaf is calculated correctly by adding previous leaves related attributes and that there is no negative balance in any leaf.

So now, we commit to a liabilities tree that includes all of our user liabilities. We also hold the total amount we commit to and have a proof for the correctness of the entire tree and calculation, without revealing the total amount of our liabilities. This allows us to maintain privacy while still proving our solvency.

\subsection{Proof Statement}

In this section, we will discuss the entire proof statement of our proof of liabilities process.

In simple terms, we've developed a commitment tree with a specific leaf structure that allows us to make the following proof statement: "I know a Merkle tree with a public root and nodes, the leaves of which contain data about my user balances. The last right leaf in this tree contains data about the total sum of my liabilities per asset."

There are two important aspects of this proof statement:

\begin{itemize}
    \item I have a Zero-Knowledge Proof (ZKP) for this tree that demonstrates that the data in the last leaf is calculated correctly and that there is no negative balance in any leaf. This ZKP proof allows me to verify the integrity and correctness of the total liabilities without having to reveal the actual amounts. \cite{whyzkpworks}
    \item My users can download the snapshot related to their balances from the website and also get the path to verify that their balances are included in the total liabilities. This allows users to independently confirm that their balances have been taken into account in the proof of liabilities.
\end{itemize}
Based on this approach, as a business, I only need to publish the Merkle root and nodes, not the leaf data. With the ZKP, I prove that my total obligations are included in the last right leaf without having to make this data public. This maintains the privacy of the total liabilities while still allowing users and auditors to verify the correctness of the proof of liabilities.

In the Proof of Solvency section, we will revisit the last right leaf and discuss how it aids us in maintaining the privacy of the total liabilities while enabling efficient and correct verification.

Note: In implementation, it is advisable to use a Merkle Sparse Tree. This reduces some complexity and makes the circuit process more efficient.

\end{multicols}

{\color{gray}\hrule}
\begin{center}
\section{Proof of Reserve}
\small{In the context of a cryptocurrency business, Proof of Reserve involves demonstrating that the business holds enough cryptocurrency assets to cover the balances it owes to its customers.}
\end{center}
{\color{gray}\hrule}
\begin{multicols}{2}

Proof of Reserve is a method by which a business can demonstrate that it holds the necessary reserves to meet its obligations. Currently, the most common approach involves revealing an address and moving funds to that address, and then proving ownership of the address by signing a message or pre-announcing the address before transferring funds.

However, there are several problems with this approach:

As a business dealing with a large number of users, maintaining numerous crypto addresses can be cumbersome. Aggregating all funds into one or a few known addresses for the proof of reserve incurs significant costs.
All of the business's addresses become publicly known, affecting security and privacy. Third parties could potentially track all transactions and discover important data that could impact the business's security.
To achieve a private proof of reserve, we have identified two potential approaches, each with their own pros and cons:

With the introduction of EIP-7503 \cite{eip7503}, we realized that the basic idea of the EIP could be used to achieve our goal. Instead of transferring funds to a burnt address and privately proving a burn, we could transfer funds to a multiplication of a point on the elliptic curve for which we know the private key of the reference point. After the transfer, we can create a proof that we're transferring funds to a multiplication of the reference point (destination point = reference-point $* g^{s}$). Since no one knows 's', our destination address remains private, and we can create a proof using the transaction root in a block that we've done that. However, there are some issues with this approach:
    \begin{itemize}
    \item a. The business still needs to transfer their funds to prove their reserve.
    \item b. In Solidity, we only have access to the last 256 block hashes, which makes proof generation difficult.
    \end{itemize}
We then came up with the idea of using the state of each blockchain instead of privately proving a specific transaction. This approach solves the problems of the previous approach but it's tricky and differs from one blockchain to another. In the following sections, we will discuss how we can use the state to create such a proof in Ethereum and then in Bitcoin.\newline

Proof of Reserve involves two key steps:

\begin{enumerate}
    \item Demonstrating ownership of an address: This step involves proving that a particular address used in the Proof of Reserve process is indeed owned by the business. This is important to ensure that the assets being accounted for in the Proof of Reserve are actually controlled by the business and not by a third party.
    \item Proving a specific balance in the owned address: Once ownership of the address is established, the next step is to prove that this address holds a specific balance of a particular asset. This confirms that the business holds the necessary reserves to meet its obligations.
\end{enumerate}
We will discuss the proof of address ownership in the Proof of Solvency section. In the following section, we will explore how we can prove that a specific address has at least a minimum balance.

\subsection{Ethereum}
Ethereum is a public blockchain protocol that uses an account-based model for its accounting system. This model is a key aspect of Ethereum's architecture and allows for a wide range of financial transactions and applications.

The protocol distinguishes between two types of accounts: externally owned accounts (EOAs) and contract accounts. EOAs are controlled by private keys and are used for simple transactions. Contract accounts, on the other hand, are governed by their internal code and are used to create smart contracts.

Ethereum also introduced the concept of a Virtual Machine (VM), specifically the Ethereum Virtual Machine (EVM). The EVM is a runtime environment that executes smart contracts on the Ethereum network. These smart contracts are self-executing contracts with the terms of the agreement directly written into lines of code.

These features make Ethereum an incredibly powerful platform for a wide range of decentralized applications, including but not limited to financial applications.

In addition to its account-based model and the Ethereum Virtual Machine (EVM), Ethereum employs a data structure called the Merkle Patricia Tree (MPT) to manage the state of the entire blockchain network. The World State, a crucial component of Ethereum's architecture, is represented by the MPT, enabling the efficient storage and retrieval of account information, balances, and smart contract data. The MPT is a key data structure that enhances the integrity and security of the Ethereum blockchain by providing a tamper-evident way to organize and update the state of the network, allowing for rapid verification and validation of transactions and contracts. This combination of the MPT and the EVM forms the backbone of Ethereum's decentralized ecosystem, facilitating the creation and execution of complex, secure, and transparent applications.\cite{yellowpaper}

\subsubsection{World State and MPT}
The World State in Ethereum is the global state of the Ethereum system, which is composed of many smaller objects known as accounts. Each account is a data structure that contains four fields: the nonce, balance, storageRoot, and codeHash. The nonce is a scalar value equal to the number of transactions sent from this address, the balance is a scalar value equal to the number of Wei owned by this address, the storageRoot is a 256-bit hash of the root node of a Merkle Patricia tree, and the codeHash is the hash of the EVM code of this account.

The Merkle Patricia Trie (MPT) is a cryptographic data structure that maps keys to values. In the context of Ethereum, the MPT is used to map addresses to account states. It is a modified version of the Patricia Trie and the Merkle tree, hence the name. The MPT has three types of nodes: leaf nodes, extension nodes, and branch nodes.

\includegraphics[width=0.9\linewidth]{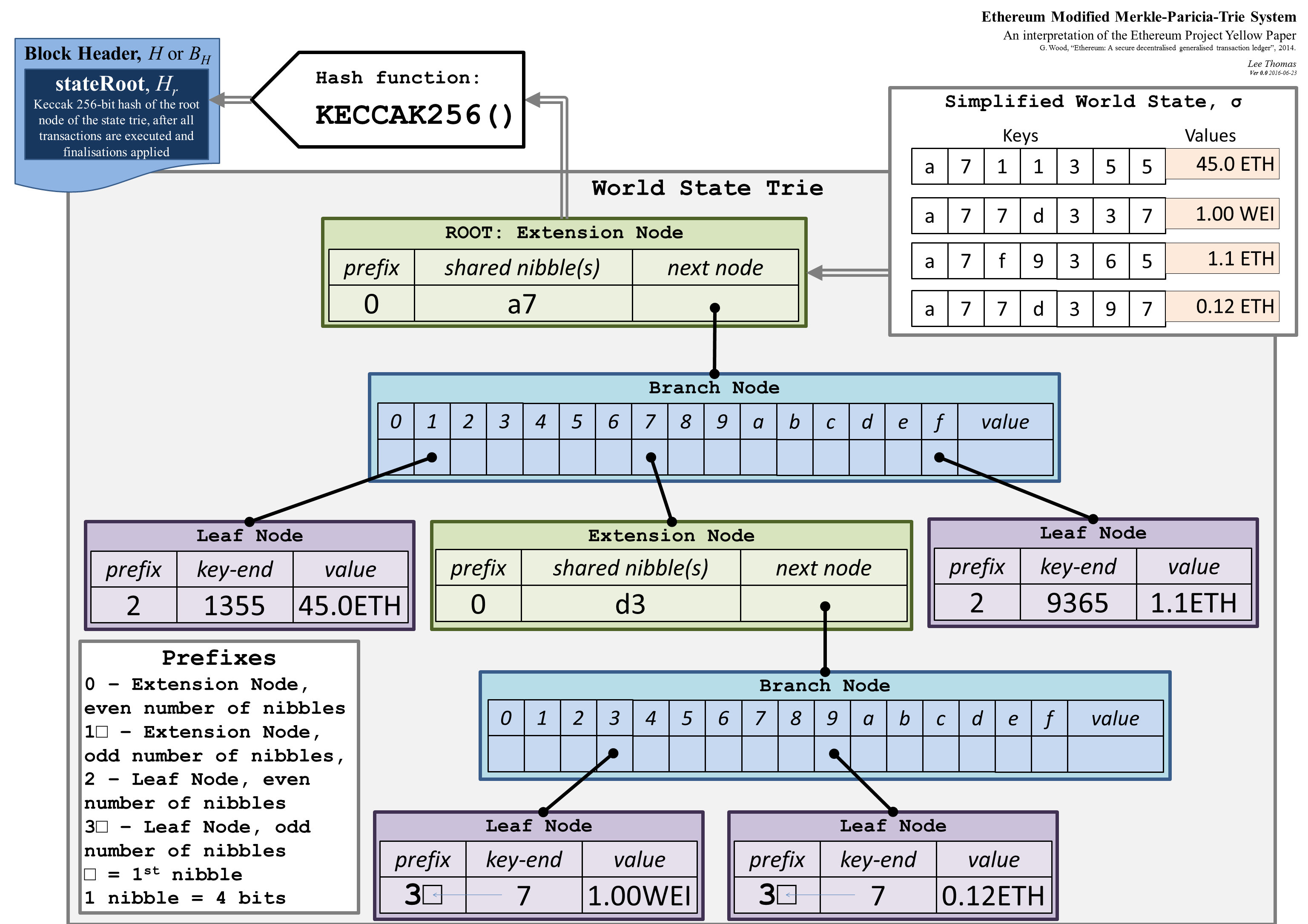}

Leaf nodes and extension nodes are similar in that they both contain a path and a value. The difference lies in what the value represents. In a leaf node, the value is the account state, whereas in an extension node, the value is the hash of the next node.
Branch nodes contain 17 items. The first 16 items point to other nodes in the trie, and the 17th item contains the value of the account state if a key ends at this node.
The root hash of the MPT, which is a cryptographic hash of all the data in the trie, is stored in the header of each block. This allows for quick and efficient verification of data.

By using the MPT, Ethereum can efficiently store the entire state of the system and quickly retrieve, update, or verify any part of it. This is crucial for maintaining the integrity and performance of the Ethereum network.

\subsubsection{GetProof Method}
The Ethereum JSON-RPC API provides an eth getProof method, a function employed by Ethereum execution nodes. This method requires an address, an array of storage keys, and a block identifier as arguments, subsequently returning an object encompassing information about the account and its storage.

The returned data includes the account's balance, nonce, storage hash, and code hash, complemented by a list of nodes (in Recursive Length Prefix, or RLP, form) that form the proof of the specified account and its storage. \cite{ethdatastructure}

This proof essentially comprises a subset of the Merkle Patricia Trie (MPT) necessary for verifying the account's data. It encompasses the MPT nodes along the path from the root to the account node, and to the storage nodes if storage keys were specified. Using this proof, the correctness of the account's data and its inclusion in the state of the specified block can be independently verified.

\includegraphics[width=0.9\linewidth]{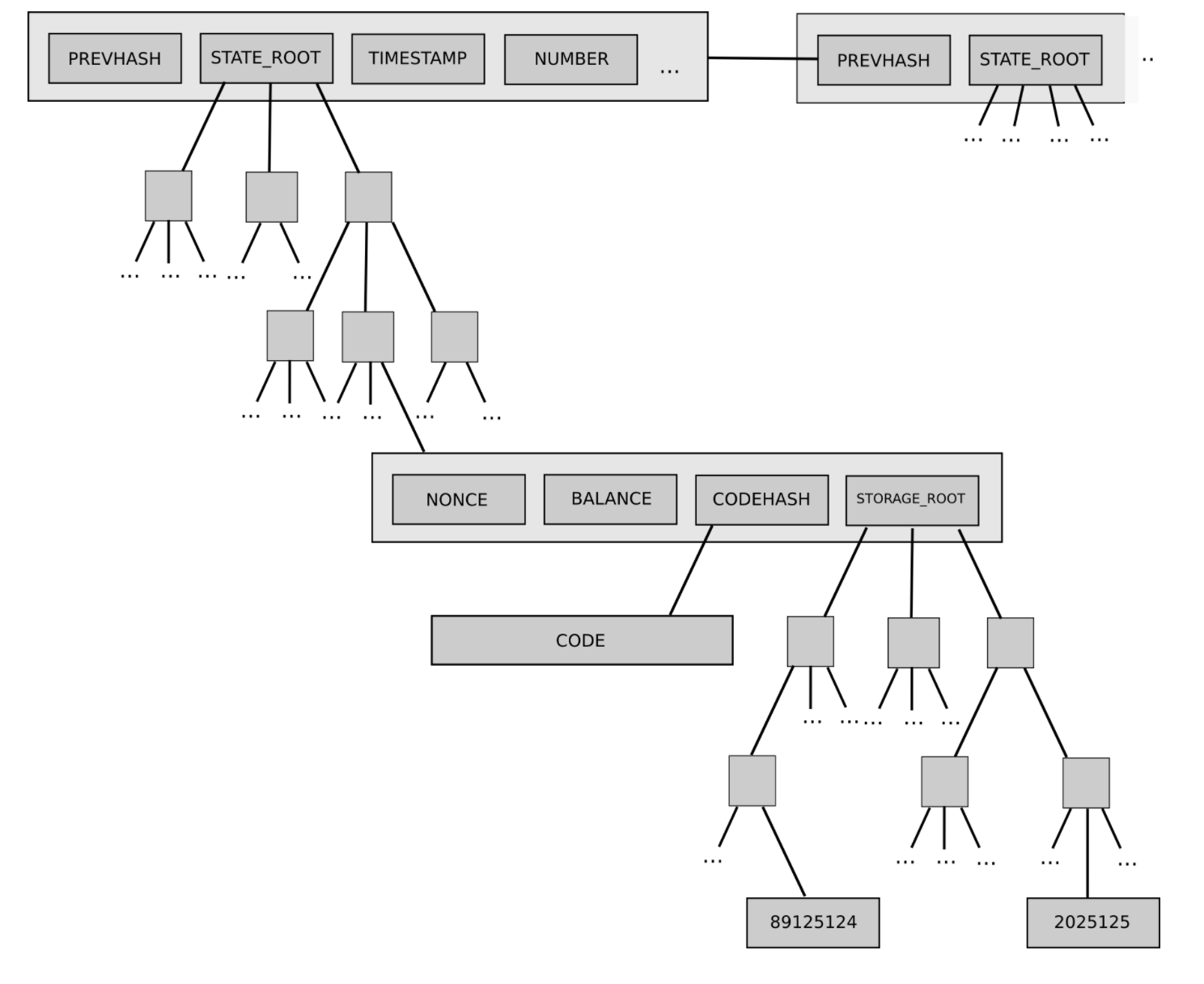}

The path within the MPT is determined by the account's address, which is employed as the key in the trie. The path to the key is a sequence of nibbles (half-bytes) derived from the key.

Proof verification involves commencing at the root node of the MPT and following the path specified by the key. Each step involves comparing the node's hash against the expected hash in the proof. If all hashes correspond and the path leads to the expected account data, the proof is validated.

This mechanism allows for the verification of an account's state at a specific block without requiring access to the entire Ethereum state, a feature that significantly enhances the scalability and efficiency of the Ethereum network.

\subsubsection{Eth Balance Proof}
Building upon the information from the previous sections, we understand that an Ethereum address is mapped to specific data in the Ethereum World State via the Merkle Patricia Trie (MPT). The balance of an account is one such data point that is mapped.

To prove that a specific address holds at least a minimum amount of Ether, we can design a Zero-Knowledge Proof (ZKP) circuit in the following manner:

Our circuit will have two public inputs:
\begin{itemize}
    \item The minimum amount
    \item The block root
\end{itemize}
And several private inputs:

\begin{itemize}
    \item MPT Proof path
    \item Account Data
    \item Block Header Data
\end{itemize}

The circuit for proving the eth minimum balance should follows this flow:

1- Hash the account data\newline 
2- Insert the hash into the MPT path\newline
3- Verify the MPT path\newline
4- Insert the calculated root as the state root alongside the other block header data\newline
5- Calculate the block hash based on the block header data\newline
6- Compare the calculated block hash with the public block root. If they match, it means we have correctly proved the data of an account in a specific block\newline
7- In the final step, check if the balance in the account data is greater than or equal to the public minimum amount.\newline

After executing all steps in our ZKP circuit, we can prove that a we know a specific address holds at least a minimum Ether balance without revealing the address itself to other people.

To implement the described flow in the ZKP circuit, certain specific functions in our zkp circuit need to be implemented. These include the Keccak hash function, Recursive Length Prefix (RLP) encoding, and MPT verification.

The Keccak hash function is a cryptographic hash function that is used in Ethereum for various purposes, including calculating the block hash and the hashes of account data and MPT nodes. Implementing this function in the circuit is crucial for verifying the MPT path and the block hash.

Recursive Length Prefix (RLP) encoding is a space-efficient object serialization scheme used in Ethereum. It's used to encode the block header data and the account data. Implementing RLP encoding in the circuit allows us to handle these data structures properly.

Lastly, MPT verification is necessary to check the validity of the proof path in the MPT. This involves following the path specified by the account's address and checking the hashes at each node against the expected hashes in the proof.

\subsubsection{ERC20 Balance Proof}
While the Ether balance proof involves verifying the state MPT for a specific address we want to prove, the ERC20 balance proof involves an additional step: verifying the contract storage for the key related to our address.

ERC20 tokens are implemented as smart contracts on the Ethereum platform. The balance of ERC20 tokens for each address is stored in the contract's storage, not in the account state as in the case of Ether. Therefore, to prove an ERC20 balance, we must access and verify the contract's storage.

The contract's storage is a separate MPT where each key-value pair is a mapping of an address to its balance. To prove a specific ERC20 token balance, we first need to generate a proof for the contract storage. This can be done using the eth getProof method by providing the key related to the specific address we want to prove. This will return a proof that can commit to the value of the token balance for that address.

\includegraphics[width=0.9\linewidth]{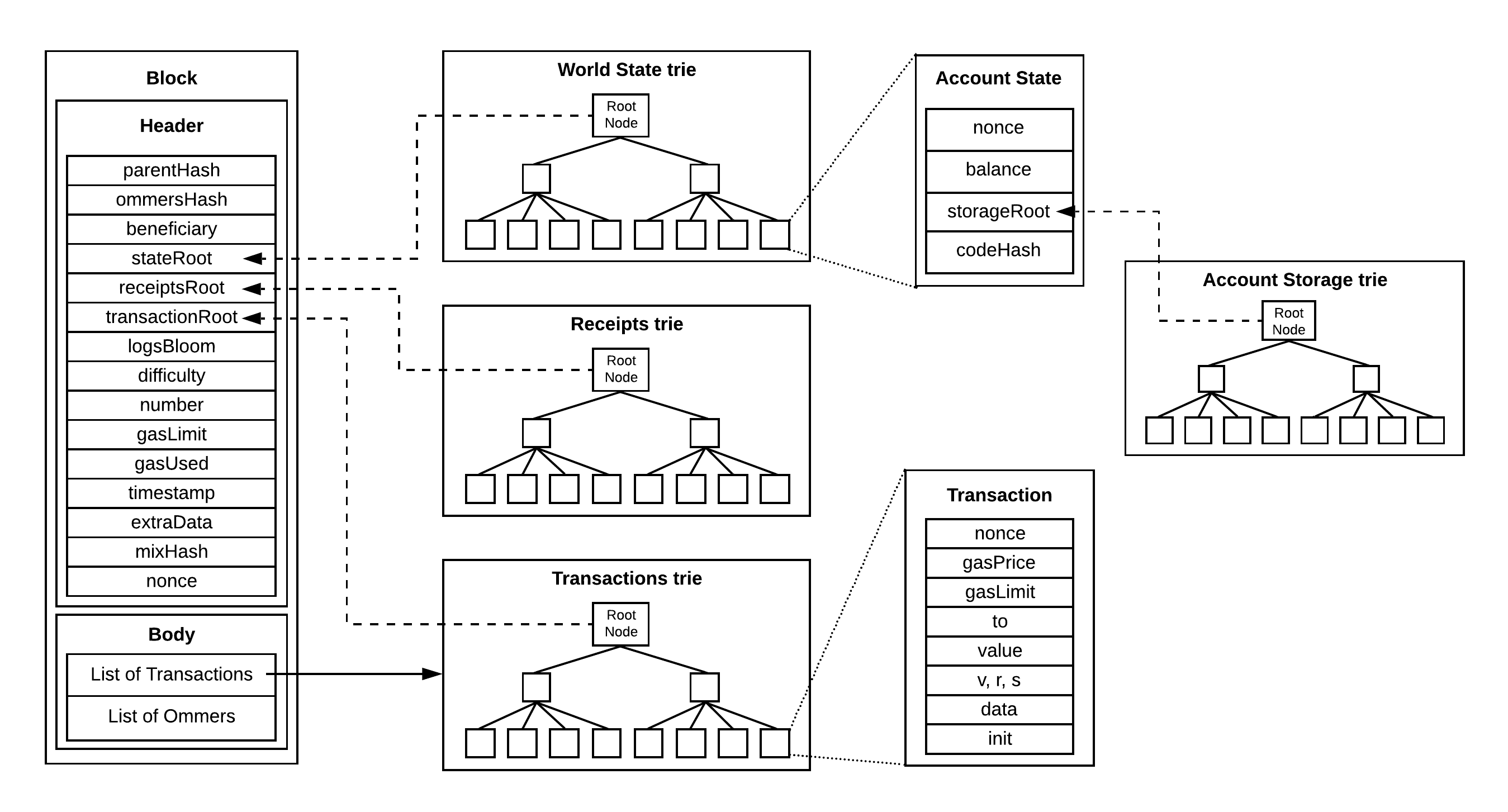}

Once we have this contract storage proof, we then verify the state MPT for the contract address, not the address we want to prove. This involves hashing the contract account data, inserting the hash into the MPT path, verifying the MPT path, and checking if the calculated balance in the contract storage proof matches the public minimum amount.

In summary, the key difference for ERC20 balance proof is the addition of verifying the contract storage via its own MPT, and using the contract address to verify the state MPT. This allows us to prove that a specific address holds at least a minimum ERC20 token balance without revealing the address itself, providing a privacy-preserving proof of reserve for ERC20 tokens.

\subsection{Bitcoin}
Bitcoin, the first decentralized cryptocurrency, was introduced in 2009 by an individual or group known as Satoshi Nakamoto. It operates on a peer-to-peer network where transactions are verified by network nodes through cryptography and recorded on a public ledger known as blockchain.

Unlike Ethereum's account-based model, Bitcoin uses an Unspent Transaction Output (UTXO) model for its accounting system. In this model, transactions output a certain number of bitcoins, which can be spent by future transactions. Each UTXO represents a chain of ownership encoded in the Bitcoin blockchain, and the sum of these UTXOs represents a user's total bitcoin balance.

Bitcoin nodes maintain a database of active UTXOs, also known as the chain state. This database is crucial for verifying new transactions, as it allows nodes to check whether the UTXOs a transaction wants to spend are indeed valid and have not been spent yet. By only keeping track of unspent bitcoins, the UTXO model simplifies transaction verification and improves the scalability of the Bitcoin network. \cite{bitcoin}

\subsubsection{UTXO and ChainStat}
At the heart of Bitcoin's operation is the Unspent Transaction Output (UTXO) model. In this model, each transaction begins with inputs that are references to previous transaction outputs and ends with outputs that specify a new Bitcoin amount and a new owner. These outputs then become the inputs of future transactions, forming a chain of ownership. Transaction types in Bitcoin, such as Pay-to-Public-Key-Hash (P2PKH) and Pay-to-Script-Hash (P2SH), govern the conditions under which these UTXOs can be spent.

\includegraphics[width=0.9\linewidth]{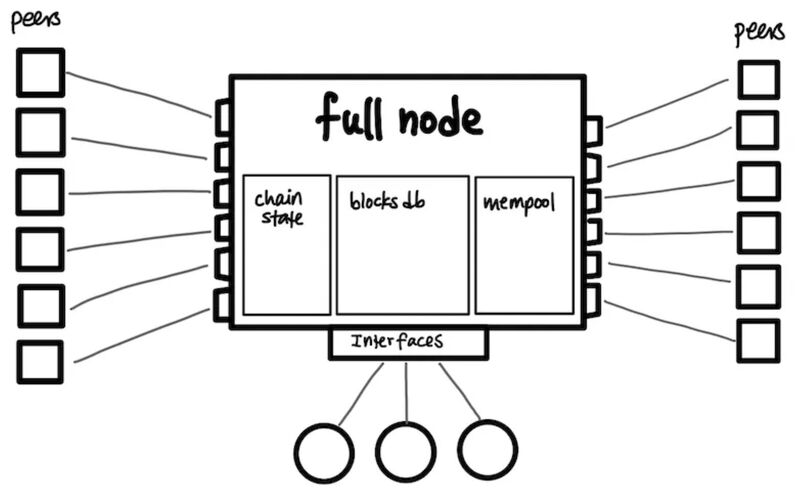}

The Chain State in Bitcoin is a key-value database that maintains a record of all active UTXOs. With the arrival of every new block, spent UTXOs are removed, and new UTXOs are added to the Chain State. This dynamic update of the Chain State ensures that it accurately reflects the current state of all transactions.

The balance of a Bitcoin address is determined by the total amount of bitcoins in the UTXOs that the address owns.

Each full node in the Bitcoin network independently maintains the Chain State. Every node computes its own Chain State by applying all transactions from the blockchain starting from the genesis block. This decentralized approach ensures that every node has a complete and independent verification of the state of all transactions, contributing to the security and robustness of the Bitcoin network.

However, this also means that the Chain State is not directly included in the block header and varies from node to node, making it challenging to create a proof of reserve that commits to a specific number of active UTXOs owned by an address. In the next section, we will discuss a solution to this challenge.

\subsubsection{Bitcoin Proof Statement}
The first approach to proving a Bitcoin balance might be to adopt the idea from EIP7503. This would involve transferring a specific amount of Bitcoin in a transaction, and then privately proving the transaction and the amount. This approach is feasible but involves operational costs, making it less than ideal.

A more efficient approach might involve the following steps:

\begin{enumerate}
    \item Announce that the platform intends to commit to its Bitcoin balances up to a specific block.
    \item Download the entire Chain State for that block from a Bitcoin node. This Chain State includes every active UTXO at the time of the block.
    \item Insert every UTXO from the Chain State as a leaf in a Merkle tree. This process converts the Chain State into a structure that allows for efficient proofs.
    \item Publish the root of the Merkle tree. Anyone can then verify the correctness of the published root with the help of a Bitcoin node and some additional code.
\end{enumerate}
With this approach, we now have a Merkle tree representing all active UTXOs at a specific block. As a business, we know which UTXOs belong to our addresses. We can then create a Zero-Knowledge Proof (ZKP) circuit to prove that we know some leaves (UTXOs) that belong to a specific address and prove their paths in the Merkle tree.

This approach allows us to commit to a specific number of active UTXOs owned by an address without revealing the address itself, providing a privacy-preserving proof of reserve for Bitcoin addresses.

\end{multicols}

{\color{gray}\hrule}
\begin{center}
\section{Proof of Solvency}
\small{Proof of Solvency is a method by which a cryptocurrency platform can demonstrate that it has sufficient funds (assets) to cover its obligations (liabilities), without revealing sensitive details about its customers or its own operations. This provides a level of transparency and trust for the platform's users, while maintaining privacy.} \cite{nicpor}
\end{center}
{\color{gray}\hrule}
\begin{multicols}{2}

Proof of solvency constitutes a crucial aspect of trust in digital asset platforms. It consists of two main components: Proof of Liabilities and Proof of Reserves. Proof of Liabilities asserts the total obligations a business has to its customers, while Proof of Reserves confirms the total assets the business holds. We have discussed various approaches to achieve private proofs for both components, and now, we aim to combine these to achieve a Private Proof of Solvency.

Let's explore the whole picture for achieving private Proof of Solvency:

Initially, a business commits to its liabilities. This involves creating a Merkle tree of all customer balances and publishing the root of this tree. the business creates a Zero-Knowledge Proof (ZKP) that attests to the correctness of the liabilities tree. The business also provides an infrastructure that allows users to verify that their liabilities have been correctly included in the tree.

In the next step the business proves its reserves using one of the methods we introduced earlier. This involves creating a proof for each of the business's addresses.

Lastly, the business integrates these proofs in a ZKP circuit to show that the total reserves are at least equal to the total liabilities. This involves comparing the total balance of specific addresses with the balance of the last right leaf in the liabilities tree. These inputs are kept private in the circuit, with the only public data being the block root and the root of the liabilities tree.

By successfully completing this process, the business achieves a private Proof of Solvency. This demonstrates that the business has sufficient reserves to cover all its liabilities without revealing any sensitive information, such as individual customer balances or business addresses. This approach not only enhances the privacy of customers and the business but also bolsters trust in the platform by providing a verifiable proof of solvency.

\end{multicols}
{\color{gray}\hrule}
\begin{center}
\section{Future Works}
\small{In the Future Works section, we will discuss potential enhancements and applications of the methods presented in this paper. We aim to inspire further exploration in the field of privacy and trust in digital asset platforms.}
\end{center}
{\color{gray}\hrule}
\begin{multicols}{2}

\subsection{Contract}
One potential area for future work involves the creation of a smart contract for storing the historical data of a business's Proof of Solvency. This contract would allow businesses to announce new rounds of Proof of Solvency, set the root of the Proof of Liabilities tree and its associated proof, and also submit proofs related to the Proof of Reserves. It would also perform a final check of solvency, ensuring that the total reserves are at least equal to the total liabilities. This could provide a transparent and verifiable record of a business's solvency over time, enhancing trust in the platform.

\subsection{Custodian Scoring Platform}
Based on the protocol introduced in this paper, we could also develop a platform for scoring custodian service providers. This platform would evaluate custodians based on their Proof of Reserves, the frequency of their solvency periods, their historical record of solvency, and other relevant factors. Such a platform could provide users with a transparent and objective way to evaluate the trustworthiness of different custodians, aiding in their decision-making process.

\subsection{User Credit}
Currently, the Proof of Solvency protocol is focused on businesses, but it could be extended to users as well. With some modifications to the protocol and collaboration from businesses, users could create private proofs that they hold at least a minimum balance in the custody of a specific business during their last submitted Proof of Solvency round. Users could also create proofs for historical balances, such as proving that they had at least a minimum balance for the last five submitted Proof of Solvency rounds by a specific business. This could provide users with a privacy-preserving way to demonstrate their financial stability and creditworthiness.

\end{multicols}
\bibliographystyle{plain}
\bibliography{main}
\end{document}